\documentclass[5p,times]{elsarticle}

\usepackage{lineno,hyperref}
\usepackage{gensymb}
\usepackage{amsmath}
\modulolinenumbers[5]

\journal{}

\begin{document}

\begin{frontmatter}

\title{First-principles study of defects at $\Sigma3$ grain boundaries
in CuGaSe$_2$}

\author[mymainaddress]{R. Saniz}
\author[mymainaddress]{J. Bekaert}
\author[mymainaddress]{B. Partoens}
\author[mysecondaryaddress]{D. Lamoen}



\address[mymainaddress]{CMT and Nanolab Center of Excellence,
Departement Fysica, Universiteit Antwerpen,
Groenenborgerlaan 171, B-2020 Antwerpen, Belgium}
\address[mysecondaryaddress]{EMAT and Nanolab Center of
Excellence, Departement Fysica, Universiteit
Antwerpen, Groenenborgerlaan 171, B-2020 Antwerpen, Belgium}

\begin{abstract}
We present a first-principles computational study of cation-Se
$\Sigma$3 (112) grain boundaries in CuGaSe$_2$. We discuss the
structure of these grain boundaries, as well as the effect of
native defects and Na impurities on their electronic properties.
The formation energies show that the defects will tend to form
preferentially at the grain boundaries, rather than in the grain
interiors.
We find that in Ga-rich growth conditions Cu vacancies as well
as Ga at Cu and Cu at Ga antisites are mainly responsible for having the equilibrium Fermi level pinned toward
the middle of the gap, resulting in carrier depletion. The
Na at Cu impurity in its +1 charge state contributes to this.
In Ga-poor growth conditions, on the other hand, the formation
energies of
Cu vacancies and Ga at Cu antisites are comparatively too high for
any significant influence on carrier density or on
the equilibrium Fermi level position. Thus,
under these conditions,
the Cu at Ga antisites give rise to a $p$-type grain boundary.
Also, their formation energy is lower than the formation energy
of Na at Cu impurities. Thus, the latter
will fail to act as a hole barrier
preventing recombination at the grain boundary, in
contrast to what occurs in CuInSe$_2$ grain boundaries.
We also discuss the effect of the defects on the electronic properties of bulk CuGaSe$_2$, which we assume reflect the properties of the grain interiors.
\end{abstract}

\begin{keyword}
CIGS; Grain boundaries; absorber layers; defects;
first-principles methods
\end{keyword}

\end{frontmatter}


\section{Introduction}

The chalcopyrite ternary and multinary compounds are materials
of prime interest because of their potential for technological
applications. The rich physical and chemical properties they
exhibit are due to a large extent to the fact that they present two or more cation sublattices, as opposed to one in binary compounds, thereby increasing their chemical degrees of freedom and the variety of dopants they can host. The applications of interest range from thermoelectrics
\cite{plirdpring2012}, through bio-sensing \cite{raevskaya2018}, to electroluminescence \cite{kim2018} and photovoltaics \cite{lee2017,spindler2019}. Research on thin-film photovoltaic cells, in particular, has seen significant advances in recent years.  Notable among these is that thin film cells based on CuGaSe$_2$ absorber layers have already breached the 10\% efficiency barrier\cite{ishizuka2019,ishizuka2014}. This is very interesting because CuGaSe$_2$ is a wide-band gap material. Indeed, although its efficiency is lower than the record-breaking efficiencies of cells based on Cu(In,Ga)Se$_2$ (CIGS) alloys \cite{jackson2016}, it presents several advantageous features because of its wider band gap \cite{ishizuka2019}. The latter results in a higher open-circuit voltage, which reduces ohmic losses due to the lower current intensity required for a given output power. Optical losses are reduced as well, as a lower number of interconnects are needed within a module. And the open-circuit loss with increasing temperature is also lower
\cite{nadenau2000}. Furthermore, its wider band gap ($E_g=1.68$ eV) makes CuGaSe$_2$ a material of choice as absorber layer in the top cell in tandem devices. Much of recent research points in this direction \cite{ishizuka2019,hu2017,wi2018,wei2018}.

Research is naturally also devoted to uncover ways to
further improve the efficiency of the CuGaSe$_2$-based
photovoltaic cells (see, e.g., reviews \citenum{ishizuka2019}
and \citenum{siebentritt2006}). Many research groups focus on film deposition and growth methods
\cite{jung2016,ullah2016,popp2017,awaah2018}, the importance of this cannot be overstated. Indeed, the efficiency of a cell is greatly impacted by the presence of defects and impurities in the various material layers in the cell and at their interfaces. Other research groups focus on the electronic properties of the materials involved and on how they are affected by defects and impurities. In this regard, important attention has been given to defects in bulk CuGaSe$_2$ \cite{spindler2019,hu2017,siebentritt2010,bekaert2015,spindler2016,islam2018} and at grain boundaries in CuGaSe$_2$ \cite{spindler2019,siebentritt2010,sadewasser2011,schmidt2012}.

In this article we present a first-principles computational study of the structural and electronic properties of native defects in CuGaSe$_2$ grain boundaries. We focus on a grain boundary of the type $\Sigma3$ \{112\}. The latter has been found to be the most common type of grain boundary in the CIGS compounds \cite{rau2009}. The grain boundaries can be Se-Se,
cation-Se, or cation-cation terminated (cation-cation terminated grain boundaries appear to be less common) \cite{abou-ras2012}. Here we are interested in studying
the effect of the above mentioned defects on the electronic properties of a cation-Se grain boundary. In a previous study of the same type of grain boundary in CuInSe$_2$, we found that such defects and impurities can have an important impact on carrier recombination and transport in polycrystalline
CuInSe$_2$ \cite{saniz2017}. Thus, it is of interest to study their properties in the case of CuGaSe$_2$ as well. For the purposes of analysis and completeness, we also consider the effect of the mentioned defects on the electronic properties of bulk CuGaSe$_2$. Moreover, we draw a comparison between our various results and experiment.

In Section 2 we introduce both the methods and the grain
boundary model used. In Section 3 we briefly review the
quantities calculated for our study and
then present and discuss our results.

\section{Methodology}

The calculations in this study were performed with the VASP
code \cite{vasp},
using the projector augmented wave (PAW) method to
describe the electron-ion interactions \cite{paw}.
The electrons treated as valence in the PAWs used are the
Cu 3d4s, Ga 3d4s4p, Se 4s4p, as well as Na 3s.
The plane wave basis set energy cutoff
was set to 500 eV. Geometry optimizations were done using
the Perdew-Burke-Ernzerhof (PBE) exchange and correlation
functional \cite{pbe}.
Forces were converged to within 0.02 eV/{\AA}, using
the VASP method based on the conjugate-gradient algorithm.
The electronic structure of the optimized geometries was
calculated using the hybrid-functional HSE06 in VASP, with
an exact-exchange fraction $\alpha=0.3089$. The latter was
found in previous work to reproduce the experimental
band gap \cite{bekaert2014}.
In the case of charged defect calculations,
a compensating homogeneous background charge density
is introduced in order to ensure charge neutrality
\cite{vasp_ch}.

\begin{figure}
\begin{center}
\includegraphics[width=0.95\hsize]{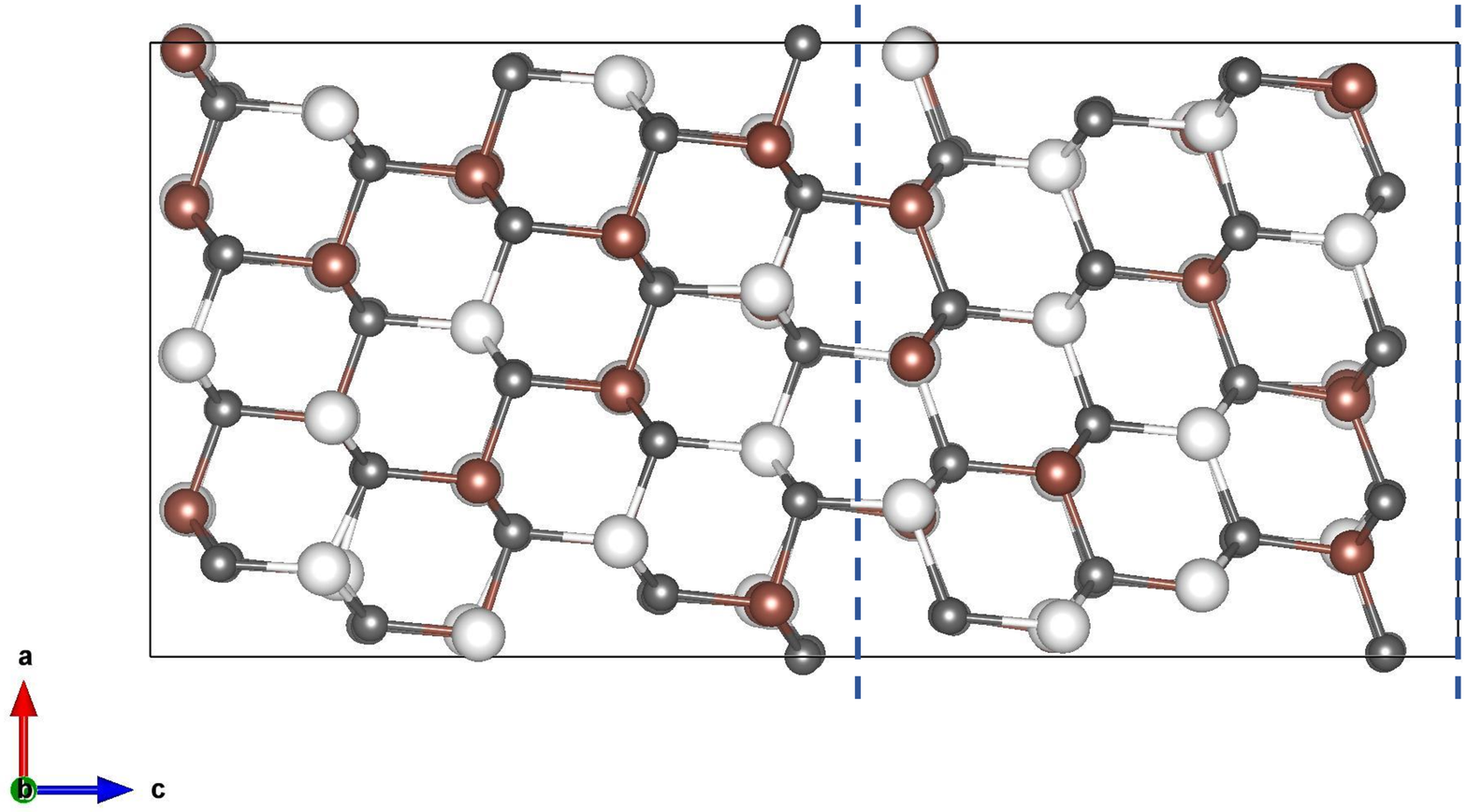}
\includegraphics[width=0.95\hsize]{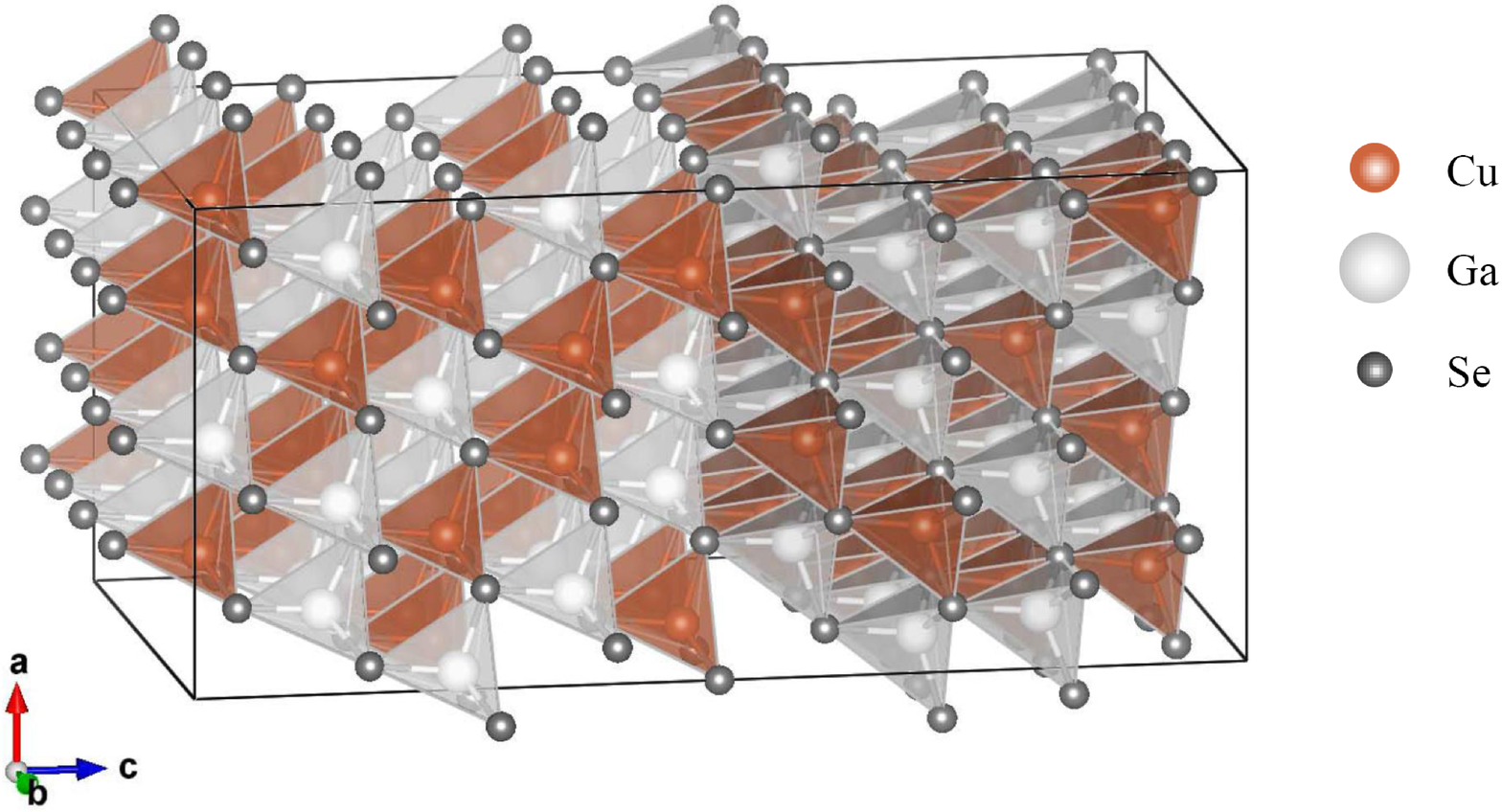}
\end{center}
\caption{\label{fig1} $\Sigma$3 (112) grain boundary model in
CuGaSe$_2$. The grain boundary plane is normal to the $c$-axis,
with its location indicated by the dashed line in the upper
panel. The `grain' on the right side can be obtained by
rotating the original structure on that side by $60 \degree$.
The tetrahedral coordination is preserved by the rotation,
but results in cation antisite pairs, with important
consequences on the properties of the GB.}
\end{figure}

The grain boundary (GB) model consists of a supercell of 288
atoms (i.e., 72 formula units), as described in more detail
further down. The dimensions of the optimized supercell are
$13.87\times16.04\times29.54$ {\AA}. We used a
$2\times 2\times 1$ {\bf k}-point mesh for all our GB calculations.
In Fig.~\ref{fig1} we present a plot of the
supercell model studied. The (112) planes are parallel
to the $ab$-plane. Because of the periodic boundary conditions,
there are two grain-boundaries. One cuts the supercell
in two along the $c$-axis, and the other is at the supercell
boundaries along the same axis. As mentioned, the grain
boundaries are cation-Se terminated, i.e., of type III in the
terminology of Abou-Ras et al. \cite{abou-ras2012}.
We note that this GB
is coherent (the four-fold coordination is conserved) and
is stoichiometry preserving. The GB can be seen as arising from
a $60 \degree$ rotation of one grain respect to the other,
i.e., around an axis perpendicular to the (112) planes
\cite{abou-ras2007}. The rotation preserves the
tetrahedral coordination of both Se and
cations, but results in cation
antisite pairs at the GB because it violates the periodicity
of CGS along the (112) planes (this is similar to what occurs
in the case of CIS).
The presence of Cu$_{\rm Ga}$-Ga$_{\rm Cu}$ antisite 
at the grain boundaries, as well as at the supercell
boundaries, leads to local octet rule violations.
Specifically, some Se atoms are coordinated not by two Cu and
two Ga atoms, but by three (four) Cu atoms and one (no) Ga atoms, or vice versa. These have an
effect on cation defect formation energies, and thus
on the relative defect concentrations and on carrier density
and transport, as we discuss in the next Section.

\section{Results and discussion}

\subsection{Calculations}

We first calculated the GB energy, defined as
\begin{equation}
\gamma=\frac{1}{2A}(E_{\rm tot}[{\rm GB}]-nE_{\rm tot}[{\rm bulk}]),
\end{equation}
where $E_{\rm tot}$[GB] ($E_{\rm tot}$[bulk]) is the total
energy of the GB supercell (of a bulk unit cell), 
$n$ is (the equivalent) number of bulk unit cells contained
in the GB supercell, and $A$ is the GB area \cite{koerner11}.
We find a GB
energy of 0.65 J/m$^2$, which is
slightly higher than in the case of CIS
(0.42 J/m$^2$). The value compares well with the energy of
very
stable twin grain boundaries in other systems. Indeed, the
$\Sigma3$ (111) grain boundary in SrTiO$_3$ and the $\Sigma7$ ($10\bar12$) grain boundary in $\alpha$-Al$_2$O$_3$ have
reported energies of 0.52 and 0.63 Jm$^{-2}$,
respectively \cite{hutt2001}. Our result is thus consistent
with the fact that this type of grain boundary is common in CGS.

The band gap reduces from the bulk value of 1.68 eV to
0.47 eV at the grain boundary.
Also, the valence band maximum (VBM) is displaced
away from the $\Gamma$ point [to (0 0.5 0) in supercell
reciprocal units], while the conductin band minimum
(CBM) remains at $\Gamma$.
Similar to CIS, the band gap narrowing obeys largely to
the presence of Cu$_{\rm Ga}$ antisites. Indeed,
this results
in several Se atoms being coordinated by 4 Cu atoms, instead
of 2. Now
the valence band is known to be of dominant Cu $d$ and Se $p$
character, with important $p$-$d$ repulsion \cite{jaffe83}.
Because 
the VBM itself is essentially of Se $p$ character
\cite{saniz2017,bekaert2014},
a stronger $p$-$d$
repulsion raises significantly the VBM with respect to the bulk.

\begin{figure}
\begin{center}
\includegraphics[width=0.95\hsize]{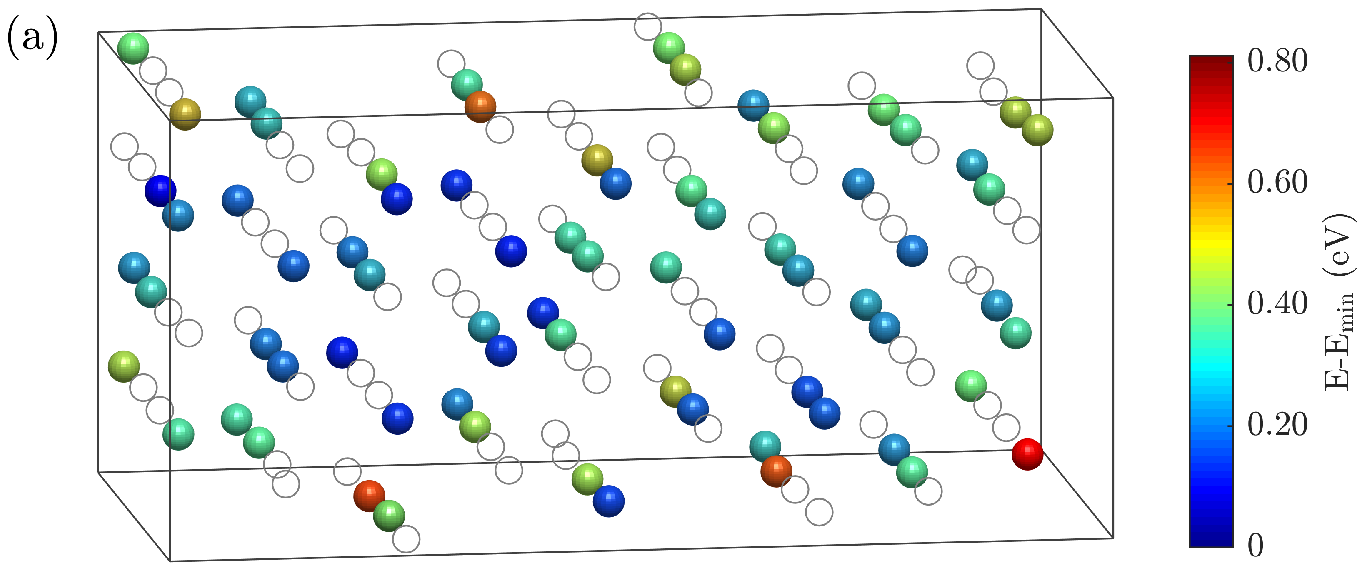}
\includegraphics[width=0.95\hsize]{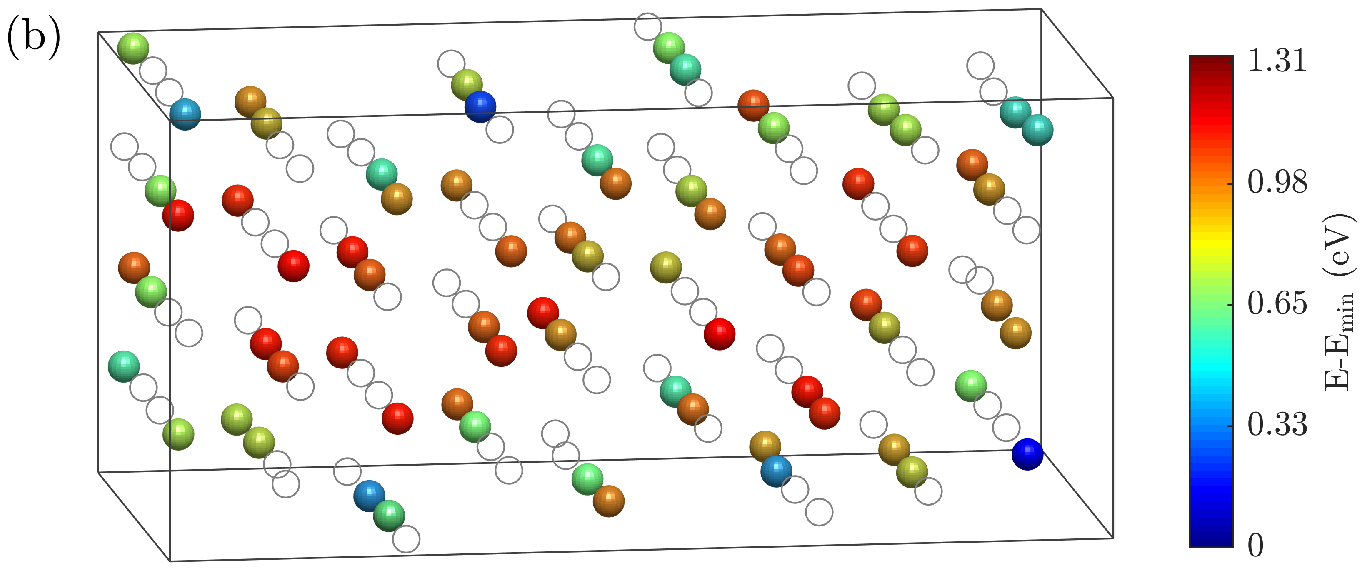}
\includegraphics[width=0.95\hsize]{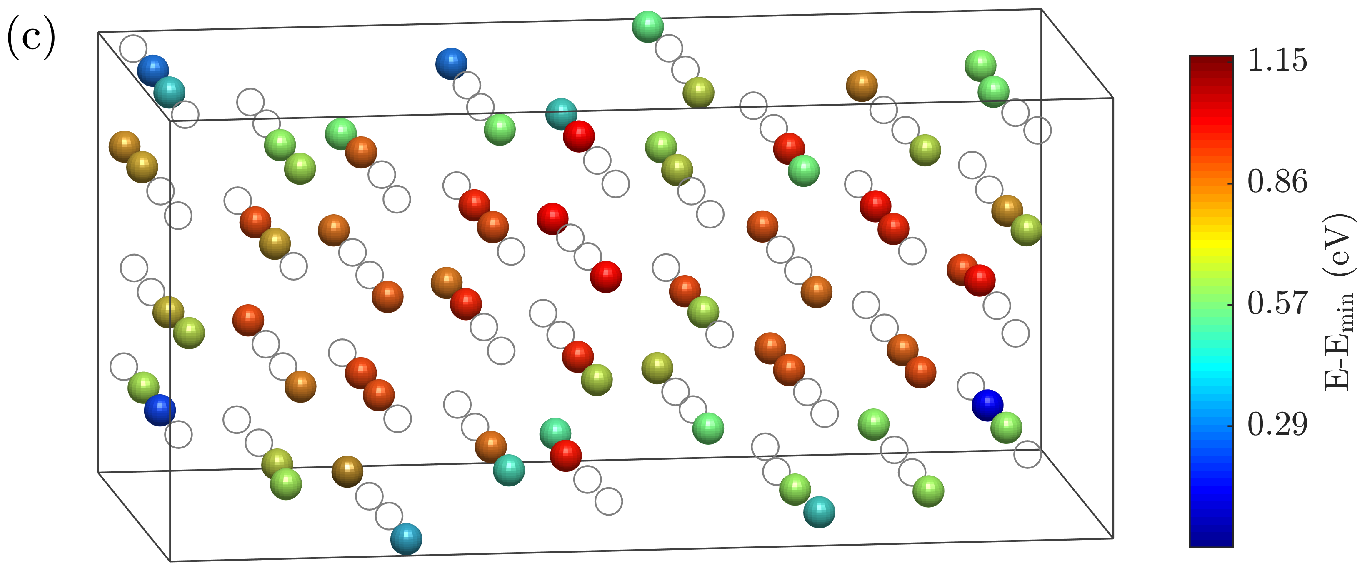}
\includegraphics[width=0.95\hsize]{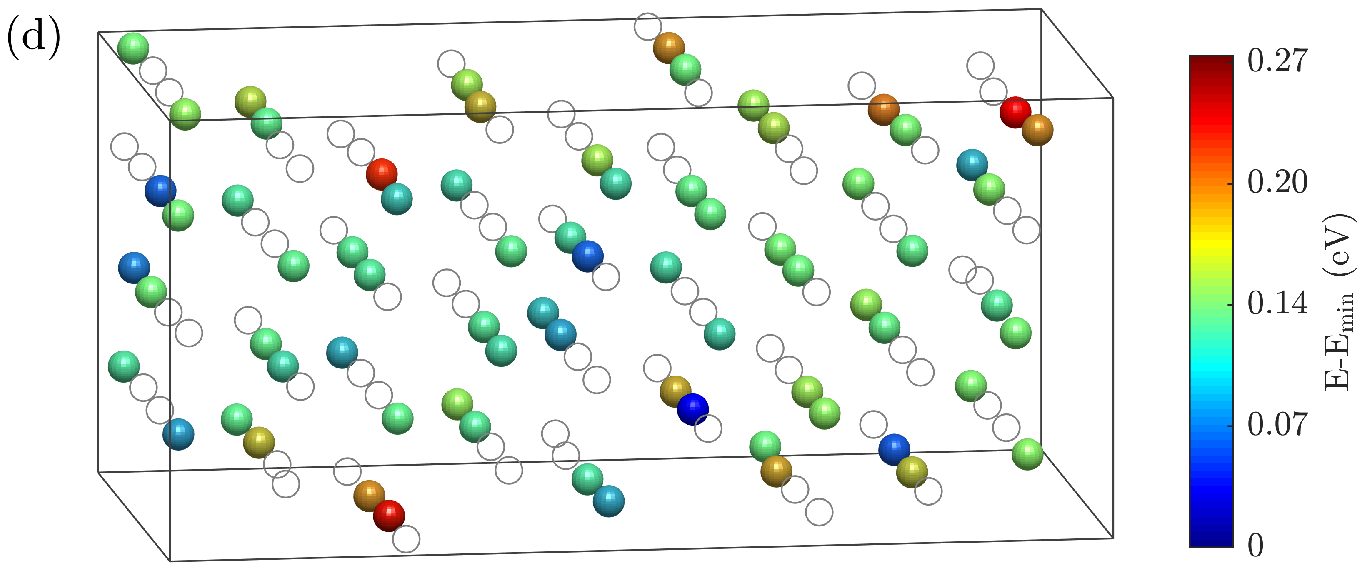}
\end{center}
\caption{\label{fig2} The color of the spheres indicate in each case whether a defect at that
location in the GB supercell (cf .Fig. 2) results in a high or low total energy,
as specified by the colorbars (energies are referred to the lowest energy in each
case). (a) corresponds to V$_{\rm Cu}$, (b) to Ga$_{\rm Cu}$, (c) to Cu$_{\rm Ga}$, and (d) to Na$_{\rm Cu}$. In
(a), (b) and (d) the gray circles indicate the positions of the Ga atoms, while in
(c) they indicate the position of the Cu atoms. For clarity, the Se atoms are not
shown.}
\end{figure}

We studied the native defects V$_{\rm Cu}$,
Ga$_{\rm Cu}$, and Cu$_{\rm Ga}$, as well as Na$_{\rm Cu}$. These appear to be the most important from the electronic properties point of view \cite{spindler2019,siebentritt2010}. Na is found in CIGS absorber layers either because of the synthesis method or because it is intentionally added \cite{kronik1998,rockett2005}. 
For a more thorough understanding of
the behavior of the defects and how they affect
the electronic properties of the GBs, we first
determined their preferred locations in the
supercell via total energy calculations, considering all 72
possible locations in each case. Figure 2 illustrates how
the total energies vary as the defect changes location in the GB
supercell (cf. Fig.2). In (a) the color of the spheres indicate
whether a V$_{\rm Cu}$ located at that position has a low or high energy.
The darker blue colors indicate a low total energy after relaxation,
while a strong red color indicates a high total energy. The
energies are referred to the lowest energy in each case. For orientation,
the gray circles indicate the position of the Ga atoms
(for simplicity, the Se atoms are not shown). (b) corresponds to
Ga$_{\rm Cu}$ and (d) to Na$_{\rm Cu}$. (c) corresponds to Cu$_{\rm Ga}$,
in which case the colored spheres indicate Ga atom positions and the
gray circles the positions of the Cu atoms. More information on
the lowest energy locations in each case is given in the
Supplementary Data file.

We then
calculated the formation energies of the defects in their
lowest energy location.
This was done
following the
standard procedure \cite{zhang1991,vandewalle2004}.
We determined the formation energies for the
defects in different
charge states, and considered Ga-rich and Ga-poor 
growth conditions.
The chemical potentials used are the same as in
Ref.~\citenum{bekaert2014}. In Fig.~\ref{fig3} we show the formation
energies as a function of the electron chemical potential,
or Fermi level, as it is commonly referred to. For a given
defect, which charge state has the lowest formation energy
depends on the Fermi level. This results in a lowest formation
energy curve consisting of segments corresponding to different
charge states, as shown by the solid lines in the plots.
Fig.~\ref{fig3}(a) corresponds to Ga-rich conditions,
and Fig.~\ref{fig3}(b) to Ga-poor conditions.
The reference potential alignment and image charge correction
were performed as in our previous CIS grain-boundaries study
\cite{saniz2017}.
For the image charge correction the experimental dielectric
constants were used \cite{syrbu1997}.
For the purposes of discussion,
we also present here the defect charge formation energies for
bulk CGS \cite{error}. The defect
formation energies in the bulk case are plotted in
Figs.~\ref{fig4}(a) and (b), the latter corresponding to
Ga-poor conditions and the former to Ga-rich 
conditions \cite{vga}. 

\subsection{Discussion}

Let us start by pointing out
that in a material with donor and acceptor defects,
in equilibrium conditions
the electron chemical potential will
actually be pinned to a value, $E_F^{\rm eq}$,
determined by overall charge neutrality \cite{sze2007}.
A rough estimate of
this energy is given by the crossing of the formation energies
of the lowest donor and acceptor lines \cite{efermieq}.
Thus, in Ga-rich
conditions, in both the GB and grain interior (the bulk
results being expected to correspond to the grain interior),
$E_F^{\rm eq}$ will
fall away from the band edges, implying carrier depletion and
very low conductivity. This is in line with has been observed
experimentally \cite{gerhard2001}.
In Ga-poor conditions, on the other hand, $E_F^{\rm eq}$ 
falls at, or near, the VBM in the GB and the bulk, respectively.
We can expect these to exhibit $p$-type conductivity in such
conditions. Experimentally, this is indeed known to be
the case \cite{schoen2000,siebentritt2003}.
Below we
examine the electronic properties of the different
defects indicated in the previous subsection
and discuss in more detail their role in the above observations.

\begin{figure}
\begin{center}
\includegraphics[width=0.7\hsize]{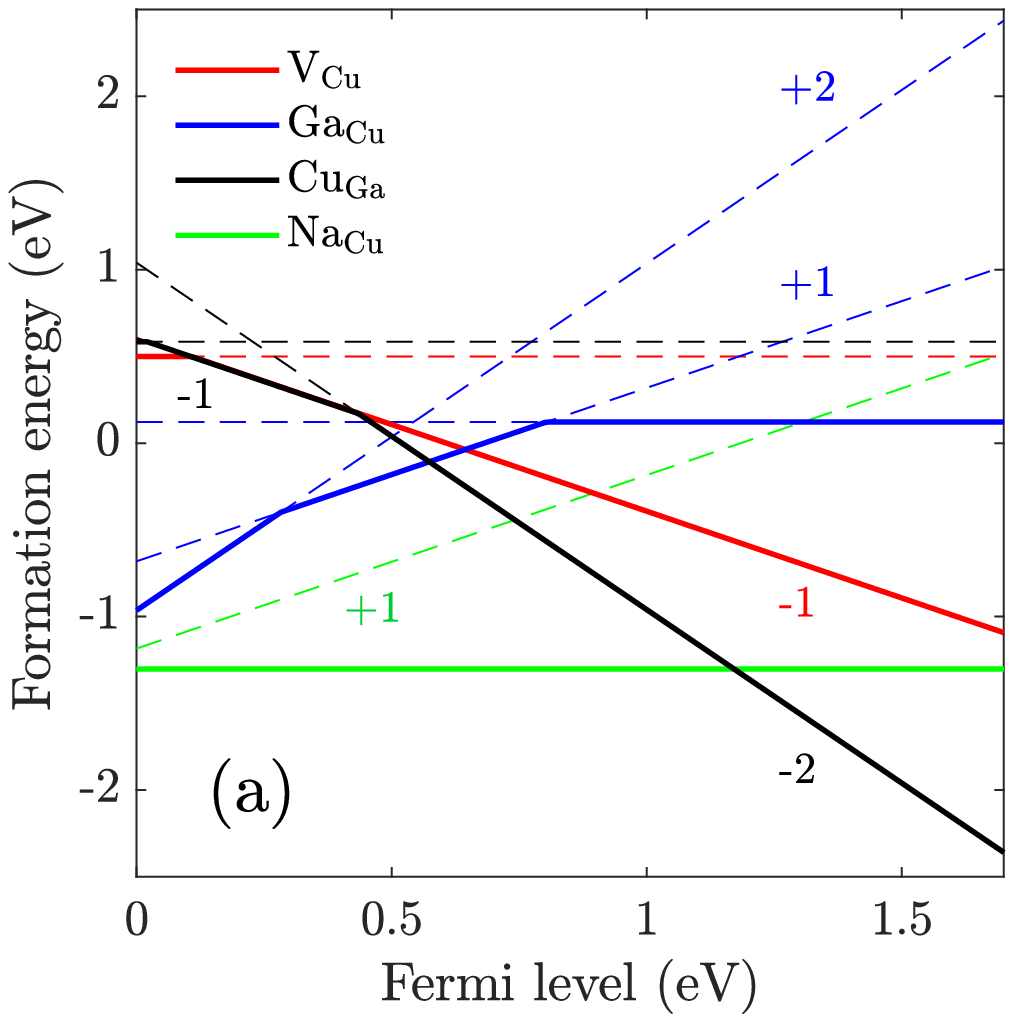}
\null\vspace{4pt}
\includegraphics[width=0.7\hsize]{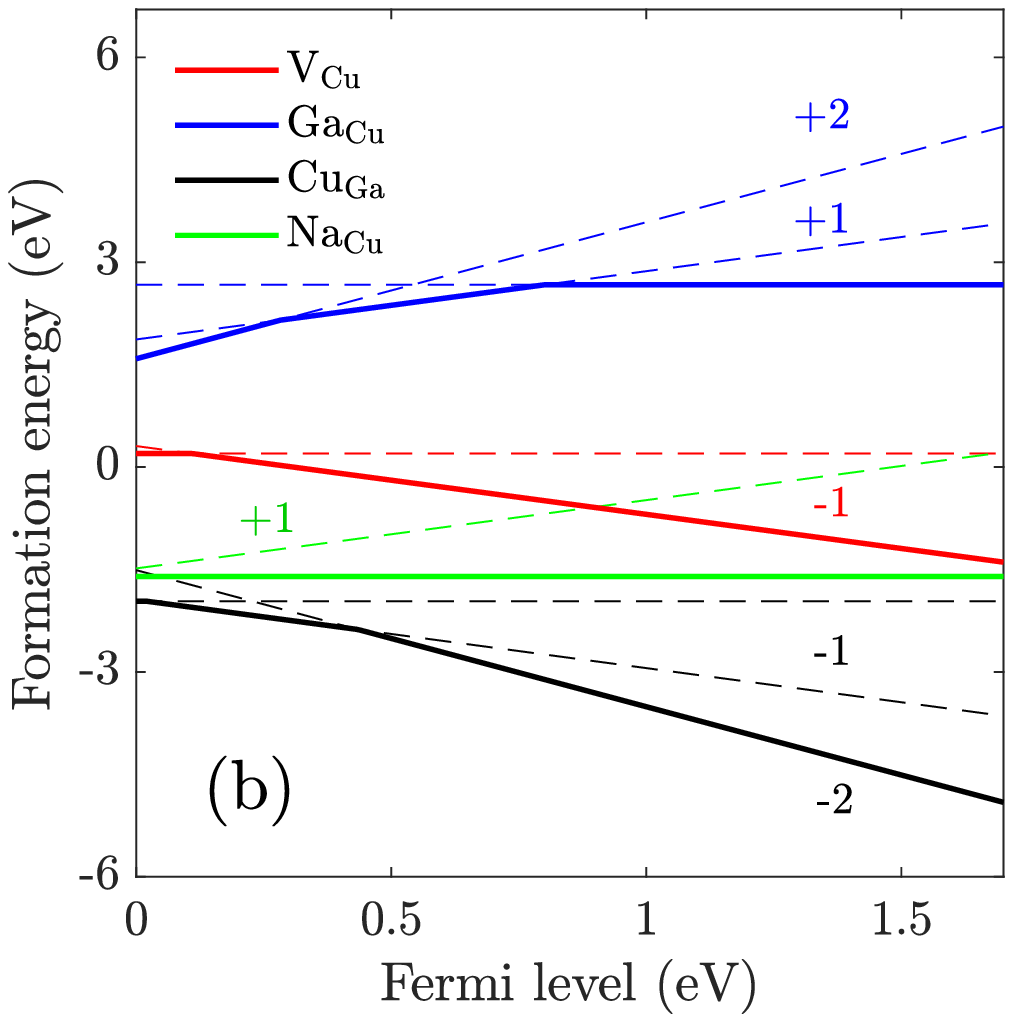}
\end{center}
\caption{\label{fig3} Formation energies of the defects in
their lowest energy location in the GB supercell,
as determined by our calculations.
(a) Ga-rich conditions; (b)
Ga-poor conditions. The solid lines indicate the preferred
charge state for each defect as a function of Fermi level.}
\end{figure}

\begin{figure}
\begin{center}
\includegraphics[width=0.7\hsize]{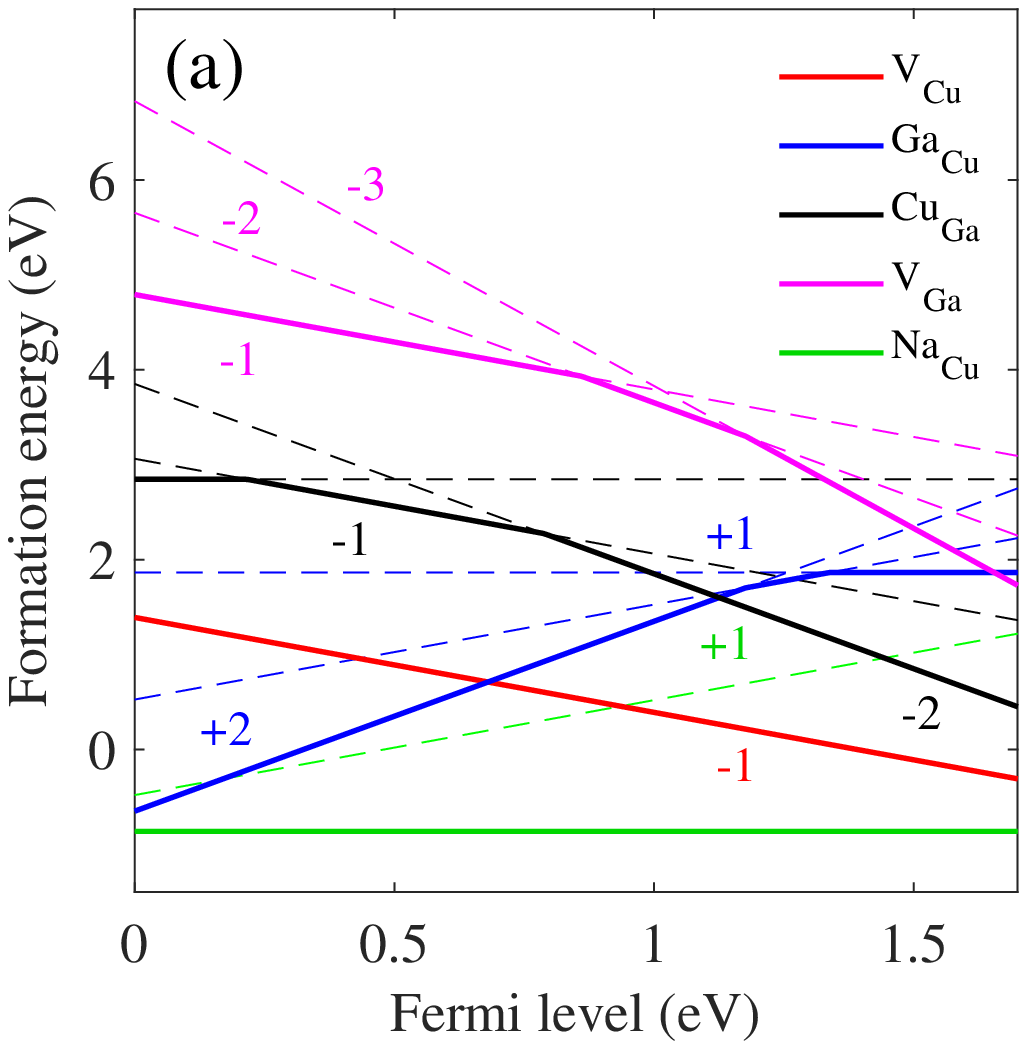}
\null\vspace{6pt}
\includegraphics[width=0.7\hsize]{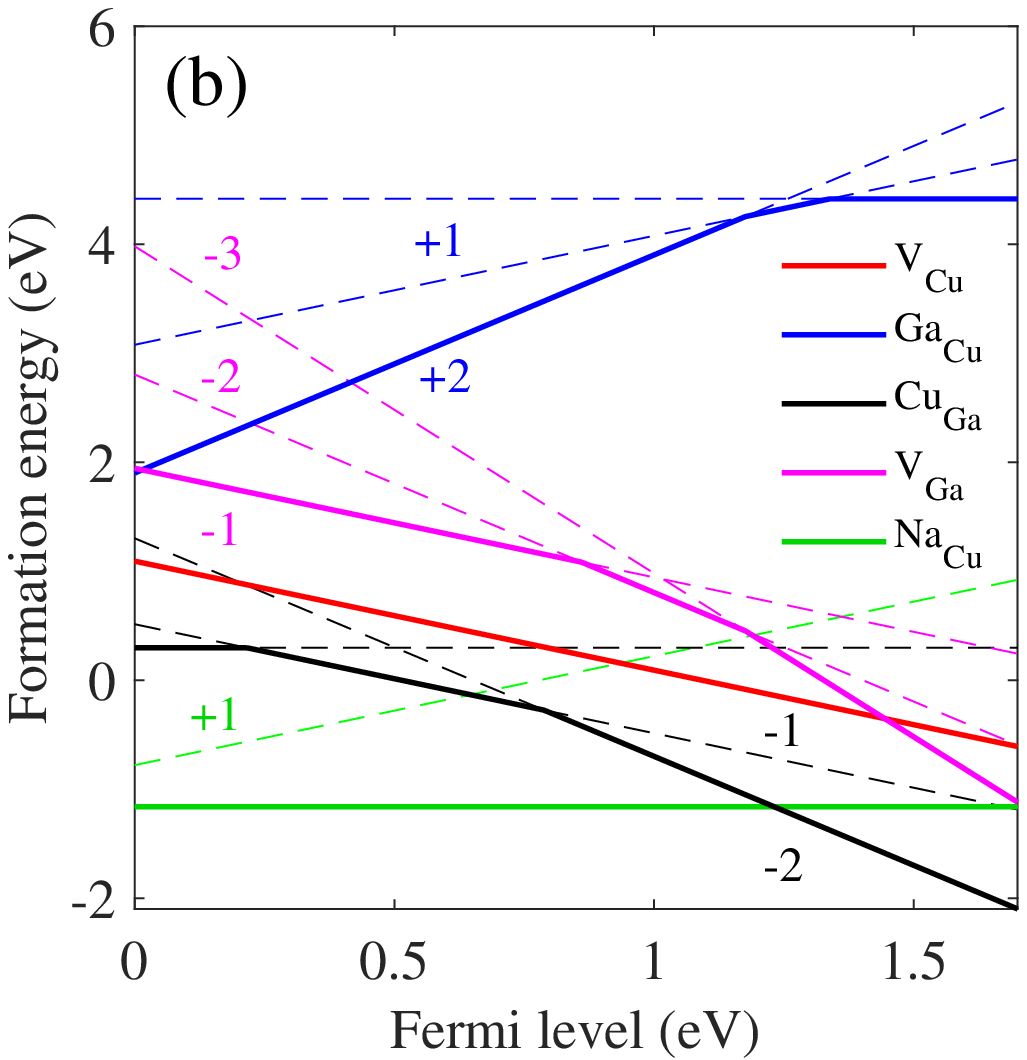}
\end{center}
\caption{\label{fig4} Formation energies of the defects in
a bulk CGS supercell. (a) Ga-rich conditions; (b)
Ga-poor conditions. As in Fig.~\ref{fig3}, the solid lines indicate the preferred
charge state for each defect as a function of Fermi level.}
\end{figure}

We consider first the Cu vacancies (V$_{\rm Cu}$). 
The lowest energy location is at the GB. But
a closer
inspection of the results leading to Fig.~\ref{fig2} reveals
that the vacancies are more
likely to form at a Cu site of a normally coordinated
Se atom, away from Se atoms coordinated by
insufficient Ga atoms.  Normally coordinated
Se atoms occur of course also within the grains. Thus, we
look at the formation 
energies in Figs.~\ref{fig3} and \ref{fig4}. These show
that the V$_{\rm Cu}$ will form preferentially at the GB,
as it has a lower formation energy by roughly 0.9 eV.
Electronically, V$_{\rm Cu}$ is clearly a
shallow acceptor in the bulk. At the GB, it may still be
considered as a shallow acceptor, as its neutral-to-negative charge transition level is $\epsilon(0|-)=106$ meV.
In Ga-rich conditions,
however, it will contribute little to any
$p$-type carrier density and transport, whether in the bulk
or in the GB. Indeed, as indicated above, $E_F^{\rm eq}$
falls near the middle of the gap. Since carrier densities
fall exponentially away from the band edges, they will be
negligible at $E_F^{\rm eq}$. In
Ga-poor conditions, the GB is $p$-type as already stated,
but the
contribution of V$_{\rm Cu}$ will be limited. This is because
Cu$_{\rm Ga}$ has a much lower formation
energy in such conditions and will be the main carrier contributor. In the bulk case, one can expect
V$_{\rm Cu}$ to be the main carrier contributing
acceptor, as the other acceptor in
this case (Cu$_{\rm Ga}$, discussed below) is somewhat
deep. Note that in experiment V$_{\rm Cu}$ is found to be
at 60 meV above the VBM \cite{spindler2019}, while our
calculations find it to be at the VBM. It is possible that
our calculation underestimates the defect level depth compared
to experiment. A source of discrepancy might be, e.g., the
charge correction scheme, which is thought to be less
appropriate for shallow defects \cite{persson2005}.

We address now Ga$_{\rm Cu}$.
According to the results in Fig.~\ref{fig2}, the
location-energy relation of the Ga$_{\rm Cu}$ antisites
is practically anti-correlated with the graph of the
V$_{\rm Cu}$ sites. This is because Ga preferably replaces
a Cu bound to a Se atom coordinated by excess Cu atoms.
Intuitively, it tends to restore the octet rule.
Octet rule violations occur more frequently at the GBs and,
in addition, the plots in Figs.~\ref{fig3} and \ref{fig4}
show that the formation of
Ga$_{\rm Cu}$ (comparing neutral charge states)
is around 1.7 eV lower at
the GB. Thus, this will be its preferred location.
Ga$_{\rm Cu}$ is a deep donor,
both in the bulk and at the GBs.
At the GB, in Ga-rich conditions, it is responsible for
pushing $E_F^{\rm eq}$ away from the VBM. The behavior is
essentially the same in the bulk. Thus, the GB and grain
interior are practically insulating in such a conditions,
which is detrimental to carrier transport and collection.
This may be one of the reasons why
a too high [Ga]/([In]+[Ga]) ratio is not beneficial to
the efficiency of Cu(In,Ga)Se$_2$ cells.
In Ga-poor conditions, the
formation energy of Ga$_{\rm Cu}$ is too high compared to
that of the other defects to have an important
effect, both at the GB and in the bulk.
Note that in the bulk
we find the donor level to be at 365 meV from
the CBM, in quite good agreement with experiment, where it is
reported to be at 400 meV below the CBM \cite{spindler2019}.

Turning to Cu$_{\rm Ga}$, Fig.~\ref{fig2} shows that its
lowest energy locations correlate strongly with those of
Ga$_{\rm Cu}$. This is not surprising since, as mentioned
before, Cu$_{\rm Ga}$ and Ga$_{\rm Cu}$ tend
to form antisite pairs at the GBs and Cu$_{\rm Ga}$ will tend
to restore the octet rule around Se atoms as
well. Furthermore, its formation energy at the
GB is close to 2 eV lower (comparing neutral states) than
in the bulk. Thus, Cu$_{\rm Ga}$ will also form preferentially
at the GBs. This is a shallow acceptor defect at the GB.
In Ga-rich
conditions it will contribute to pin $E_F^{\rm eq}$ far
from the band edges.
In Ga-poor conditions, on the other hand, it is by far the leading acceptor
and has the lowest formation energy overall, giving rise to
a $p$-type GB. This is consistent with has been observed in 
polycrystalline CuGaSe$_2$ grown under Cu-excess
\cite{schuler2002}. In the bulk Cu$_{\rm Ga}$ is a relatively
deep acceptor. In Ga-rich conditions it will
play a limited role, due to its relatively high formation 
energy. In Ga-poor conditions, it will contribute only
secondarily to $p$-conductivity, compared to V$_{\rm Cu}$.
We find that the
Cu$_{\rm Ga}$ transition levels in the bulk are
$\epsilon(0|-)=216$ meV and $\epsilon(-|-2)=790$ meV,
with respect to the VBM. These can be compared with the
Cu$_{\rm Ga}$ defect levels observed in experiment in CuGaSe$_2$
\cite{spindler2019}. An acceptor level is observed at 60 meV
above the VBM, which is assigned to a $(0|-)$ transition,
and another transition at $\sim700$ meV above the VBM,
which is tentatively assigned to a $(-|-2)$ transition.
While the theoretical and experimental $(0|-)$ transition level values are quite off of each other (again, this is in
principle a shallow level), the deep
$(-|-2)$ transition level values are in good agreement.
This supports the intepretation given to the latter in
Ref.~\citenum{spindler2019}.

We briefly point out that our present results for V$_{\rm Ga}$ in bulk CuGaSe$_2$ support previous findings \cite{pohl2013}. As seen in Fig.~\ref{fig4}, V$_{\rm Ga}$ is a shallow acceptor in the $-1$ charge state, with a high formation energy even in Ga-poor growth conditions. Our value close to 2 eV in those conditions ($\Delta\mu_{\rm Ga}=-3$ eV ) is comparable to the value of ~2.5 eV in Ref.~\citenum{pohl2013}, calculated in somewhat less Ga-poor growth conditions ($\Delta\mu_{\rm Ga}=-2.17$ eV). Due to this, V$_{\rm Ga}$ does not have a significant effect on the electronic properties of CuGaSe$_2$. This is expected to remain so in GBs, and we did not study V$_{\rm Ga}$ in GBs.

Finally, we consider the case of Na$_{\rm Cu}$ impurities. The analysis of the results in Fig.~\ref{fig2} indicates that these occur more commonly at the GBs. Also, the formation
energy of Na$_{\rm Cu}$ (neutral state) is lower at the GB by
$\sim$0.45 eV. At the GBs, in Ga-rich conditions,
although its ground state is neutral, it will still contribute
to compensate the acceptor defects because the formation energy
of its +1 charge state is sufficiently low in the lower half
of the band gap. In
Ga-poor conditions, it will tend to play the same role, but
less efficiently, as its formation energy is somewhat higher
than the donor in that case (Cu$_{\rm Ga}$). This is in
contrast with the case of CIS, where at the GBs
Na$_{\rm Cu}$ impurities clearly tend to passivate any donors
and can even contribute to type inversion \cite{saniz2017}.
It must be pointed out, however, that theoretical calculations
suggest that Na$_{\rm Cu}$ can give rise
to a neutral hole-barrier at GBs in CIGS compounds, resulting from its
lack of $d$-states compared to Cu \cite{saniz2017,persson2003}, an effect indeed observed in
CuGaSe$_2$ \cite{siebentritt2006b}.
Thus, Na$_{\rm Cu}$ at the
GB can be expected to be overall slightly beneficial.
In the bulk case, in Ga-rich conditions Na$_{\rm Cu}$ will
not alter significantly
the $n$-type {vs.} $p$-type carrier balance, so it will remain
carrier depleted. In Ga-poor conditions, Na$_{\rm Cu}$ will
tend to passivate any acceptor. This is detrimental, of
course, to the intended $p$-type conductivity of the bulk.
But its negative impact will be mitigated, as Na is expected
to segregate to the GBs, where it can have a beneficial effect.

\section{Conclusions and outlook}

Our first-principles computational study of cation-Se
$\Sigma$3 (112) grain boundaries in CuGaSe$_2$ shows
that the native defects V$_{\rm Cu}$, Ga$_{\rm Cu}$, and
 Cu$_{\rm Ga}$ and the impurity
Na$_{\rm Cu}$ have a formation energy at the GBs that is lower
than in the grain interiors. Thus, they will
have a higher concentration at the GBs. With respect to their
effect on
electronic properties, we find that in Ga-rich growth conditions
V$_{\rm Cu}$ and the antisites Ga$_{\rm Cu}$ and Cu$_{\rm Ga}$ 
are mainly responsible for having $E_F^{\rm eq}$ pinned toward
the middle of the gap, resulting in carrier depletion. But
in Ga-poor growth conditions, the formation energies of
V$_{\rm Cu}$ and Ga$_{\rm Cu}$ are high and do not influence
significantly the carrier density or the position
of $E_F^{\rm eq}$, while Cu$_{\rm Ga}$ locates the latter
near the VBM and gives rise to a $p$-type grain boundary.
Furthermore,  Na$_{\rm Cu}$ will be
in its neutral state, and while in Ga-rich
conditions its
$+1$ state still contributes to pin $E_F^{\rm eq}$ toward the
middle of the gap, in Ga-poor conditions its effect is weak
and will fail to passivate the Cu$_{\rm Ga}$ defects and
prevent recombination at the grain boundary, in
contrast with its effect in CuInSe$_2$ GBs. However, it is
still expected to act as a neutral hole barrier due to its
band bending effect at the GBs.
As regards
bulk CuGaSe$_2$, we find that V$_{\rm Cu}$ is a
shallow donor, while in experiment it is found to be 60 meV
above the VBM. On the other hand, we find that Ga$_{\rm Cu}$
is a deep donor, 365 meV below the CBM, which compares well
with the experimental value of 400 meV. Also, for
Cu$_{\rm Ga}$ we find a deep $(-|-2)$ transition level at
790 meV above the VBM, supporting the interpretation
given to the level $\sim700$ meV above the VBM observed
in experiment.
Finally, we find that Na$_{\rm Cu}$ tends to have a
detrimental effect on the carrier density, both in Ga-rich
and Ga-poor conditions, but greatly mitigated because it
will typically segregate to the GBs and have a low density in
the grain interiors.

There are several paths that future work can take.
Here we have considered a set of defects that affect
critically the electronic properties of GBs in CuGaSe$_2$, and
thus of the absorber layer as a whole. But there are other types
of defects that can be considered as well. In CIGS absorber layers,
for instance, depending
to an extent on the synthesis method, potassium and heavier alkali
impurities can be present \cite{jackson2016}, or carbon 
\cite{hibberd2010} and oxygen \cite{niu2017}. Moreover, such
defects can be present at the interfaces between the absorber
layer and the substrate or buffer layers, with important effect
on carrier passivation/transport.
The effects of such impurities have been studied 
computationally mainly
in CuInSe$_2$ so far \cite{xiao2014,sahoo2018,music2019}. It
will be important that future work considers them in CuGaSe$_2$
as well.

\section*{Acknowledgments}

We acknowledge the financial support of FWO-Vlaanderen
through project G.0150.13
The computational resources
and services used in this work were provided by the VSC (Flemish
Supercomputer Center) and the HPC infrastructure of the University
of Antwerp (CalcUA), both funded by FWO-Vlaanderen and the
Flemish Government-department EWI.


\begin{thebibliography}{99}
\bibitem{vga} Note that Ga vacancies (V$_{\rm Ga}$) are also considered further down in the case of the bulk. This is for completeness and for comparison with previous work reported in Ref.~\citenum{bekaert2014}.
\end{thebibliography}
\end{document}